\pdfoutput=1
\documentclass[11pt,twoside,a4paper,cmspaper,final,collab]{cms-tdr}
\def\svnVersion{ef47a51}\def\svnDate{2022/01/10}\def\cmsCernNoTag{CERN-EP-2021-134}\def\cmsCernDate{\today}\def\cmsMessage{Published in Physical Review Letters as \href{http://dx.doi.org/10.1103/PhysRevLett.128.081805}{\doi{10.1103/PhysRevLett.128.081805}.}}
\begin{document}\cmsNoteHeader{HIG-20-015}

\newlength\cmsTabSkip\setlength{\cmsTabSkip}{1ex}

\newcommand{\mtautau}{\ensuremath{m_{\PGt\PGt}}\xspace}
\newcommand{\pttau}{\ensuremath{\pt^{\tauh}}\xspace}
\newcommand{\pth}{\ensuremath{\pt^{\PH}}\xspace}
\newcommand{\njets}{\ensuremath{N_\text{jets}}\xspace}
\newcommand{\mjj}{\ensuremath{m_\mathrm{jj}}\xspace}
\newcommand{\ljpt}{\ensuremath{\pt^{\mathrm{j}_1}}\xspace}
\newcommand\mynobreakpar{\par\nobreak\@afterheading} 
\ifthenelse{\boolean{cms@external}}{\providecommand{\cmsLeft}{upper\xspace}}{\providecommand{\cmsLeft}{upper left\xspace}}
\ifthenelse{\boolean{cms@external}}{\providecommand{\cmsRight}{lower\xspace}}{\providecommand{\cmsRight}{lower\xspace}} 
\ifthenelse{\boolean{cms@external}}{\providecommand{\cmsMiddle}{center\xspace}}{\providecommand{\cmsMiddle}{upper right\xspace}} 

\cmsNoteHeader{HIG-20-015} 
\title{Measurement of the inclusive and differential Higgs boson production cross sections in the decay mode to a pair of \texorpdfstring{$\PGt$}{tau} leptons in \texorpdfstring{$\Pp\Pp$}{pp} collisions at \texorpdfstring{$\sqrt{s} = 13\TeV$}{sqrt(s) = 13 TeV}}

\author*[cern]{The CMS Collaboration}

\date{\today}

\abstract{
Measurements of the inclusive and differential fiducial cross sections of the Higgs boson are presented, using the \PGt lepton decay channel. The differential cross sections are measured as functions of the Higgs boson transverse momentum, jet multiplicity, and transverse momentum of the leading jet in the event, if any. The analysis is performed using proton-proton collision data collected with the CMS detector at the LHC at a center-of-mass energy of 13\TeV and corresponding to an integrated luminosity of 138\fbinv. These are the first differential measurements of the Higgs boson cross section in the final state of two \PGt leptons. In final states with a large jet multiplicity or with a Lorentz-boosted Higgs boson, these measurements constitute a significant improvement over measurements performed in other final states.
}

\hypersetup{
pdfauthor={CMS Collaboration},
pdftitle={Measurement of the inclusive and differential Higgs boson production cross sections in the decay mode to a pair of tau leptons at sqrt(s) = 13 TeV},
pdfsubject={CMS},
pdfkeywords={CMS,  Higgs, taus, fermion couplings, differential cross section, fiducial cross section}}

\maketitle 

Measuring differential production cross sections of the Higgs boson could eventually highlight the contribution 
of beyond-the-standard-model physics to the Higgs boson couplings~\cite{Grazzini:2016paz,Bishara:2016jga}, \eg, 
by the observation of deviations from the standard model (SM) in the Higgs boson transverse momentum (\pt) distribution, 
predicted with high accuracy at next-to-next-to-leading order (NNLO) precision~\cite{Alioli:2019qzz}. 
Such measurements are also powerful probes of the SM predictions, in particular of the higher-order corrections in perturbation theory, and could help improve event modelling.  

Differential cross sections of Higgs boson production have been measured in the $\PGg\PGg$, $\PZ\PZ$, 
$\PWp\PWm$, and $\PQb\PAQb$ decay channels for various sets of observables, by the ATLAS and CMS Collaborations at the CERN LHC at 
center-of-mass energies of 7, 8, and 13\TeV~\cite{Khachatryan:2016vnn,Aaboud:2018ezd,Sirunyan:2018sgc,Sirunyan:2018kta,Sirunyan:2020tzo,ATLAS:2020wny,CMS:2021ugl}. 
The $\PH\to\PGt^+\PGt^-$ decay channel~\cite{CMS-HIG-16-043,Aaboud:2018pen} can also contribute to differential measurements of the Higgs boson production, 
providing complementary information with other decay modes. It is competitive in parts 
of the phase space where small production cross sections are compensated by a relatively large branching fraction $\mathcal{B}(\PH\to\PGt^+\PGt^-)=6.2\%$~\cite{PDG2020}; this 
is particularly the case for high jet multiplicities ($\njets$) and large 
Lorentz boosts of the Higgs boson. This Letter presents the first differential fiducial measurements of the 
Higgs boson production cross section using its decays to a pair of \PGt leptons. The Higgs boson cross section 
is measured as functions of its transverse momentum ($\pth$), $\njets$, 
and the leading jet \pt ($\ljpt$), using data collected by the 
CMS experiment in proton-proton ($\Pp\Pp$) collisions at a center-of-mass energy of 13\TeV between 2016 and 2018, 
corresponding to an integrated luminosity of 138\fbinv. 
A measurement of the inclusive fiducial 
Higgs boson cross section is also presented, in a phase space complementary to those studied with other final states. 

The central feature of the CMS apparatus is a superconducting solenoid of 6\unit{m} internal diameter, providing a magnetic field of 3.8\unit{T}. Within the solenoid volume are a silicon pixel and strip tracker, a lead tungstate crystal electromagnetic calorimeter, and a brass and scintillator hadron calorimeter, each composed of a barrel and two endcap sections. Forward calorimeters extend the pseudorapidity coverage provided by the barrel and endcap detectors. Muons are detected in gaseous detectors embedded in the steel flux-return yoke outside the solenoid.
A more detailed description of the CMS detector, together with a definition of the
coordinate system used and the relevant kinematic variables, can be found in
Ref.~\cite{Chatrchyan:2008zzk}.

Simulated events with Higgs bosons are generated for the different production modes 
(gluon fusion, vector boson fusion, and productions in association with a vector boson, {\PW} or {\PZ}, or with top quarks) 
at next-to-leading order (NLO) precision in perturbative quantum chromodynamics (QCD), including finite 
quark mass effects, with
the \POWHEG 2.0~\cite{Nason:2004rx,Frixione:2007vw, Alioli:2010xd, Alioli:2010xa, Alioli:2008tz} generator.
The distributions of $\pth$ and $\njets$ in the gluon fusion production simulation are corrected to match the predictions 
of the $\textsc{nnlops}$ generator~\cite{Hamilton:2013fea,Hamilton:2015nsa}.
The Higgs boson mass is assumed to be 125.38\GeV~\cite{Sirunyan:2020xwk}.

The \MGvATNLO 2.2.2 (2.4.2) event generator~\cite{Alwall:2014hca} is used to simulate the Drell--Yan process at
leading order with the MLM jet matching and merging scheme~\cite{Alwall:2007fs} for the simulation of data taken in 2016 (2017 and 2018).
It is also used to model the diboson production at NLO in $\alpS$, whereas \POWHEG
2.0 and 1.0 are used for {\ttbar} and single top quark production, respectively.
Single top quark production in the $t$-channel 
and diboson events are normalized to their cross sections at NLO precision or higher~\cite{Campbell:2011bn,Gehrmann:2014fva}.
Drell--Yan events, as well as {\ttbar} events and single top quark production in the $\PQt\PW$-channel, 
are normalized 
to their cross sections at NNLO precision~\cite{Melnikov:2006kv,Czakon:2011xx}. 
The generators are interfaced with \PYTHIA 8.212~\cite{Sjostrand:2014zea} to model
the parton showering and fragmentation, as well as the decay of the \PGt leptons.
The \PYTHIA tunes {CUETP8M1} and {CUETP8M4}~\cite{CMS-PAS-TOP-16-021} are used in simulation corresponding to the 2016 data-taking conditions, 
and the {CP5} tune~\cite{CMS:2019csb} is used for 2017 and 2018 simulations. The parton density function (PDF) set is NNPDF 3.0 for 2016 simulations, 
and NNPDF 3.1 for 2017 and 2018 simulations~\cite{Ball:2011uy,Ball:2013hta,Ball:2017nwa}. 
Additional proton-proton interactions per bunch crossing, called pileup, are added to the simulations with the profile observed in data. 
Simulated events are processed through a \GEANTfour~\cite{Agostinelli:2002hh} simulation of the CMS detector.

The particle-flow (PF) algorithm~\cite{Sirunyan:2017ulk} is used to reconstruct the events on the basis of 
information from the different CMS subdetectors. 
Muons are reconstructed from tracks and hits in the tracker and muon systems~\cite{CMS:2018rym,CMS:2019ied}.
Electrons are reconstructed from tracks in the tracking system,
and calorimeter deposits, and identified with a multivariate discriminant described in Ref.~\cite{CMS:2020uim}. 
The relative isolation of electrons (muons) is calculated on the basis of the \pt of tracks in a cone of $\Delta R = \sqrt{\smash[b]{(\Delta \eta)^2 + (\Delta \phi)^2}} < 0.3$ (0.4) centered 
on the lepton track, corrected for charged and neutral pileup contributions; it is required to be less than 0.15. 
Jets are clustered from PF candidates using the anti-$\kt$ \FASTJET algorithm with distance parameter $R$ of 0.4~\cite{Cacciari:2008gp,CMS-PAS-JME-16-003}, 
requiring $\pt>30\GeV$ and $\abs{\eta}<4.7$. Jet energy corrections are applied on an event-by-event 
basis~\cite{Khachatryan:2016kdb,Sirunyan:2017ulk,CMS-DP-2020-019}. 
In events collected in 2017, jets with $\pt < 50\GeV$ and $2.65 < \abs\eta < 3.14$ are discarded to eliminate spurious jets caused by  
detector noise. 
Hadronic jets originating from {\PQb} quarks are tagged with the medium working point of 
the \textsc{DeepCSV} algorithm~\cite{Sirunyan:2017ezt}.
The hadrons-plus-strips algorithm~\cite{Sirunyan:2018pgf}, which 
combines 1 or 3 tracks with energy deposits in the calorimeters, 
is used to reconstruct \PGt leptons decaying hadronically, 
denoted as $\tauh$.  
Deep neural network discriminants are used to reduce the fraction of quark and gluon jets, electrons, and muons 
misidentified as $\tauh$ candidates~\cite{CMS-DP-2019-033}.
All particles reconstructed in the event are used to determine the missing transverse momentum,
\ptvecmiss, which is defined as the negative vectorial sum of the transverse momenta of
all PF candidates originating from the primary$\Pp\Pp$ interaction vertex, which is the vertex with the largest value of summed physics-object $\pt^2$~\cite{CMS:2019ctu}. It is adjusted for the effect of jet energy corrections.
Corrections to the $\ptvecmiss$ are applied to reduce the mismodeling of the simulated
$\PZ+\text{jets}$ and Higgs boson samples~\cite{CMS-HIG-16-043}.

Events are selected in four final states: 
$\Pe\PGm$, $\Pe\tauh$, $\PGm\tauh$, and $\tauh\tauh$. In the $\Pe\PGm$ final state,  
a combination of triggers requiring an electron and a muon is used, and in the $\tauh\tauh$ final state, the triggers 
require the presence of two isolated $\tauh$ candidates. In the $\Pe\tauh$ ($\PGm\tauh$) final state, the events are selected 
with a trigger that relies on the presence of a single electron (muon) with \pt above 25--32 (22--24)\GeV, or a trigger that requires 
both an electron with $\pt>24\GeV$ and a $\tauh$ candidate with $\pt>20$--27\GeV (a muon 
with $\pt>19$--20\GeV and a $\tauh$ candidate with $\pt>27$--30\GeV) if the lepton \pt is too low to satisfy the single-lepton trigger thresholds. In the 
$\tauh\tauh$ final state the triggers select two $\tauh$ candidates with $\pt>35$--40\GeV. The thresholds 
depend on the data-taking year. The offline event selection criteria
are given in Table~\ref{tab:selection}, where the symbol $\mT$ denotes the invariant mass between two objects in the transverse plane. In the $\Pe\PGm$, $\Pe\tauh$, and $\PGm\tauh$ 
final states, the small fraction of events without a reconstructed jet with $\pt>30\GeV$ and with $\Delta R$ between the visible decay products of the two \PGt leptons below 2, is vetoed because 
of the difficulty in accurately estimating the backgrounds in this particular topology. 
In the $\tauh\tauh$ final state, all events are required to contain at least one jet. 
This requirement significantly reduces the QCD multijet background, while it does not affect 
the signal acceptance significantly since the Higgs bosons need to be boosted for their decay products to 
pass the high-\pt trigger thresholds. All events with a jet tagged as originating 
from a bottom quark are discarded in the $\Pe\PGm$, $\Pe\tauh$, and $\PGm\tauh$  final states, 
where the {\ttbar} background would otherwise be consequential. 

\begin{table}
\centering
\topcaption{Event selection criteria. The \pt ranges are related to different triggers used during different data-taking periods. In events collected in 2016 in the $\PGm\tauh$ channel, $\tauh$ candidates with $0.2<\abs{\eta}<0.3$ are discarded because of a significantly larger misidentification rate of muons as $\tauh$ objects.}\label{tab:selection}
\begin{scotch}{lcccc}
 & $\Pe\PGm$ & $\Pe\tauh$ & $\PGm\tauh$ & $\tauh\tauh$ \\
\hline
$\pt^\Pe$ (\GeVns) & $>$15/24 & $>$25--26 & \NA & \NA \\
$\abs{\eta^\Pe}$ & $<$2.4 & $<$2.1 & \NA & \NA \\
$\pt^\PGm$ (\GeVns) & $>$24/15 & \NA & $>$20--21 & \NA \\
$\abs{\eta^\PGm}$ & $<$2.4 & \NA & $<$2.1 & \NA \\
$\pttau$ (\GeVns) & \NA & $>$30 & $>$30 & $>$40 \\
$\abs{\eta^{\tauh}}$ & \NA & $<$2.3 & $<$2.3 & $<$2.1 \\
$\mT(\Pe/\PGm,\ptvecmiss)$ (\GeVns) & \NA & $<$50 & $<$50 & \NA \\
$\mT(\Pe+\PGm,\ptvecmiss)$ (\GeVns) & $<$60 & \NA & \NA & \NA \\
$\njets$ & \NA & \NA & \NA & $>$0 \\
\end{scotch}
\end{table}

The fiducial region is defined to be as close as possible to the reconstructed event 
selection. All variables used in the definition of the fiducial region are calculated 
at the generator level after parton showering and hadronization, and the electrons and muons are ``dressed''
in that the lepton momentum 
includes the momenta of photons radiated within a cone of $\Delta R < 0.1$ centered on the lepton. 
In the $\Pe\tauh$ ($\PGm\tauh$) final state, the electron (muon) is required to have \pt
above 25 (20)\GeV and $\abs{\eta}<2.1$, while the $\tauh$ candidate must have a visible \pt
greater than 30\GeV and visible $\abs{\eta}<2.3$. Here, the term visible refers to the 
kinematic variables constructed from the momenta of the visible decay products of the \PGt leptons, excluding the invisible neutrinos. In addition, the transverse mass $\mT(\Pe/\PGm,\ptvecmiss)$ must be 
less than 50\GeV. In the $\tauh\tauh$ final state, the visible \pt  of 
both $\tauh$ must exceed 40\GeV, while their visible $\abs{\eta}$ must be within 2.1, and there must be at least one jet with $\pt>30\GeV$. 
In the $\Pe\PGm$ final state, the leading (subleading) lepton must have 
$\pt>24$ (15)\GeV, both leptons must have $\abs{\eta}<2.4$, and the $\mT$ of the dilepton system and $\ptvecmiss$
must be below 60\GeV to remove the overlap with the $\PH\to\PW\PW$ measurement~\cite{Sirunyan:2020tzo}.
Decays of the Higgs boson other than $\PH\to\PGt\PGt$ are considered to be outside the fiducial region.
About 95\% of $\PH\to\PGt\PGt$ events passing the reconstructed event selection belong to the fiducial 
region as estimated from simulation.
The SM prediction for the Higgs boson
cross section in this fiducial region is $408\pm27\unit{fb}$, using the inclusive cross sections and branching fractions 
in Refs.~\cite{deFlorian:2016spz,Denner:2011mq,Ball:2011mu} and the fiducial acceptance from the NLO predictions of the \POWHEG 2.0 generator 
with corrections 
from the $\textsc{nnlops}$ generator for the gluon fusion production mechanism. In particular, the gluon fusion simulation is normalized to the cross section 
computed at next-to-NNLO (N$^3$LO) QCD accuracy and NLO electroweak precision. Events outside the fiducial region are treated as backgrounds in the
measurement and are normalized to their SM expectations. This treatment is chosen because most nonfiducial events 
correspond to Higgs boson decays to a pair of $\PW$ bosons, especially in the $\Pe\PGm$ final state, for which 
the differential distributions have been measured to be compatible with the SM expectation~\cite{Sirunyan:2020tzo}. 

The di-{\PGt} background, mainly composed of $\PZ\to\PGt\PGt$, leptonically decaying $\ttbar$, and diboson processes, 
is modelled with an ``embedded sample''~\cite{Sirunyan:2019drn}, where muons from dimuon events in data are replaced 
with simulated \PGt leptons. The background with jets misidentified as $\tauh$ candidates is estimated from data 
with a so-called ``misidentification rate method''~\cite{Sirunyan:2018qio}. The probability for loosely isolated jets to be misidentified as $\tauh$ is measured in control regions 
enriched in QCD multijet, $\PW+\textrm{jets}$, or $\ttbar$ events, as a function of $\pttau$, for different $\njets$, and separately 
in the barrel and endcaps of the detector. Differences between processes, $\njets$, and detector region, are typically of the order of 15, 10, and 10\%, respectively. 
The misidentification probabilities are corrected on an event-by-event basis depending on the \pt of the other \PGt lepton in the event, $\pth$, and $\ljpt$, 
with multiplicative corrections ranging 0.5--1.2 for each variable. 
The reconstructed variable $\pth$ is evaluated as the
vectorial \pt sum of the visible decay products of the \PGt leptons and $\ptvecmiss$, multiplied with

a correction factor that is measured in signal simulation and depends on this same vectorial sum to make it an unbiased estimator of the generated $\pth$. The correction factor
reaches a plateau between 1.05 and 1.10 at high $\pth$ values, and is significantly below 1.0 at low $\pth$ values. For
events with $\pth>350\GeV$ at the generator level, the reconstructed $\pth$ resolution is better than 10\%, whereas it is worse 
than 30\% for $\pth<45\GeV$.

The misidentification probabilities as a function of the \pt of the other \PGt, $\pth$ and $\ljpt$ were measured after the initial tau \pt misidentification measurement due to the large number of
variables impacting the misidentification probabilities. These corrections are determined by a comparison of data-to-prediction distributions in the aforementioned control regions.
Additionally, corrections for the selection criteria that differ 
between the signal and control regions, such as the same-sign charge requirement for the \PGt leptons in the QCD-enriched region and the high $\mT$ requirement in the $\PW$-enriched region, 
are introduced, and depend on the reconstructed di-{\PGt} mass, $\mtautau$. They are typically close to 1.0 but can reach up to 1.2 in parts of the 
phase space. In the 
$\Pe\tauh$ and $\PGm\tauh$ final states, the overall misidentification  
rate is a weighted average of the corrected misidentification rates measured for the different types of processes. The 
weights are proportional to the expected fraction of each process with respect to the total background, 
determined event-by-event as a function of $\njets$ and $\mtautau$, using simulations for the $\PW+\textrm{jets}$ 
and {\ttbar} backgrounds. In the $\tauh\tauh$ final state, the misidentification probabilities are measured only in the dominant QCD multijet background. They are used 
to reweight events where the leading $\tauh$ candidate fails the $\tauh$ identification criteria.
The very small contribution of events where 
only the subleading $\tauh$ is a jet but the leading $\tauh$ is genuine is estimated from simulation. 

The background with jets misidentified as electrons or muons in the $\Pe\PGm$ final state, essentially events from QCD multijet, $\PW+\textrm{jets}$, and 
semi-leptonically decaying $\ttbar$ production, is estimated from 
data events where the electron and the muon have same sign, reweighted with an extrapolation factor that 
depends on $\njets$ and $\Delta R(\Pe,\PGm)$. 
Other backgrounds are estimated from simulation and scaled to their theoretical cross sections. 

To increase the signal sensitivity without introducing a strong model dependence, events are classified in different
categories depending on $\pttau$.
In the $\Pe\tauh$ and $\PGm\tauh$ final states, the
categories are defined with the following requirements: $30<\pttau<50$, $50<\pttau<70$, and
$\pttau>70\GeV$. In the $\tauh\tauh$ channel the requirements are based on the subleading $\tauh$ candidate because
the misidentification probability decreases with $\pttau$:
$40<\pttau<50$, $50<\pttau<70$, and
$\pttau>70\GeV$. No categorization is introduced in the $\Pe\PGm$ channel because the signal-to-background ratio
does not significantly increase with the lepton \pt.

Systematic uncertainties are associated with the triggering and reconstruction of the different objects selected in 
the analysis and they amount to typically 2--3\% in the efficiency and 0.5--3.0\% in the energy scale, per object.    
Uncertainties in the small misidentification rates of electrons and muons as $\tauh$ candidates range between 5 and 40\% depending on the decay mode 
and $\eta$, while the uncertainty in the momentum scale for these objects is up to 6\%. Similar uncertainties, 
partially correlated, are considered for the objects in the embedded samples~\cite{Sirunyan:2019drn}. 
Uncertainties in the jet momentum scales and $\ptvecmiss$ measurement are evaluated event-by-event. 
The uncertainty in the {\PQb} tagging reaches up to 10\% for processes with heavy-flavor jets. 

Uncertainties of 2.0, 4.2, 5.0, and 5.0\% are used for the predicted cross sections of the Drell--Yan, $\ttbar$, 
single top quark, and diboson productions, respectively~\cite{Melnikov:2006kv,Czakon:2011xx,Campbell:2011bn,Gehrmann:2014fva}. The $\PZ\to\PGt\PGt$ process yield, which is estimated with 
embedded samples, has an uncertainty of 4\% to account for the dimuon trigger used to select the initial events 
in data before the muons are replaced with \PGt leptons. 
Additionally, an uncertainty of 10\% is assigned to the normalization of embedded events without any jet in the 
$\Pe\tauh$ and $\PGm\tauh$ final states, to cover for a potential mismodeling introduced by the
 $\mT(\Pe/\PGm,\ptvecmiss)$ selection criterion.  

Several sources of uncertainty are taken into 
account for the estimate of the background with jets misidentified as $\tauh$ candidates: statistical uncertainties in 
the misidentification rate measurement as a function of $\pttau$; systematic uncertainties in the description of other variables 
($\ljpt$, $\pt^{\Pe/\PGm}$, and $\pth$), as determined from closure tests; systematic uncertainties in the extrapolation 
between the regions where the misidentification rates are measured and the signal region; and systematic uncertainties to cover 
for a finer granularity of some variables in the signal region, \eg, signal regions with 2, 3, or 4 jets while 
the misidentification rates are measured inclusively for $\njets\geq2$. In particular, the last source of uncertainty 
includes a 5\% uncertainty in the yield of the reducible background in each bin of $\njets$. Events with misidentified jets in the highest 
$\pttau$ categories also have a yield uncertainty in the range of 5--10\%, depending on the final state. This avoids propagating constraints from the 
low-$\pttau$ categories under the assumption that the \pt dependence of the misidentification probabilities is linear.  

Statistical uncertainties in the number of simulated events in the signal region or observed event yields in the control regions 
are considered in all bins of the distributions. 
The uncertainty in the integrated luminosity for the combined 2016--2018 period is 1.6\%, while individual years have uncertainties in the range 1.2--2.5\%, with partial correlations between data-taking years~\cite{Sirunyan:2021qkt,CMS-PAS-LUM-17-004,CMS-PAS-LUM-18-002}

For the signal, uncertainties from missing higher-order corrections in the perturbative QCD expansion 
are estimated by varying the renormalization and factorization scales by factors of two. 
In the case of the gluon fusion production, the uncertainty scheme proposed in Ref.~\cite{deFlorian:2016spz} is used. 
For the signal in the fiducial region, the uncertainties are implemented in such a way that they do not modify the fiducial cross sections in any of the generator-level bins before the selection considering the shape effect only. 
The uncertainties can, however, modify the normalization of the Higgs boson events outside of the fiducial region since the cross section for these events is normalized to the SM expectation. 
The fraction of the Higgs boson events in this region is less than 3 and 8\% in the $\tauh\tauh$ and $\PGm\tauh$ final states, respectively.

In each category, two-dimensional distributions of $\mtautau$, reconstructed with a simplified 
matrix element algorithm~\cite{Bianchini:2014vza} with a resolution around 20\%, and 
of the variable considered for the differential measurement ($\pth$, $\njets$, or $\ljpt$) are 
built. In practice, this is equivalent to making $\mtautau$ distributions in different bins of the other observable. 
At the generator level, $\pth$, $\njets$, and $\ljpt$ are evaluated
 with a \textsc{Rivet} implementation~\cite{Bierlich:2019rhm} of the simplified template cross sections  
scheme~\cite{deFlorian:2016spz}, where jets with $\pt>30\GeV$ are formed from clusters of final-state particles from the primary vertex,
excluding the decay products of the Higgs boson. 
Signal events from one generator-level bin contribute to multiple reconstruction-level bins. 
By performing one simultaneous fit
over all reconstruction-level bins, the signal strength modifiers of the different generator-level
observable bins, modeled as freely floating parameters of interest, can be determined using all the selected events. This
simultaneous fit is equivalent to a signal extraction in the reconstruction-level bins and its
unfolding into generator-level bins, performed in a single step.
The signal strengths per observable range are assumed fully correlated among final states since similar phase spaces are selected with 
the fiducial region definitions.  
This unfolding procedure can be sensitive to statistical fluctuations in the observed distributions and to 
small variations in the response matrix, and  
a Tikhonov regularization of the unfolded distribution is performed by 
adding to the likelihood function a multiplicative penalty term~\cite{Tikhonov:1963,Hocker:1995kb}. 
Regularization reduces statistical fluctuations and unphysical solutions, but it can lead to undercoverage 
of the uncertainty intervals and introduce systematic biases, which, in this Letter, are negligible with respect to the systematic and statistical uncertainties. These 
effects are controlled by optimizing the strength of the regularization term with the minimum global 
correlation coefficient~\cite{Schmitt:2012kp}. The optimum regularization factor is 1.85 (1.35 and 2.35) 
for the $\pth$ ($\njets$ and $\ljpt$, respectively) measurement.

\begin{figure}
  \centering
  \includegraphics[width=0.4\textwidth]{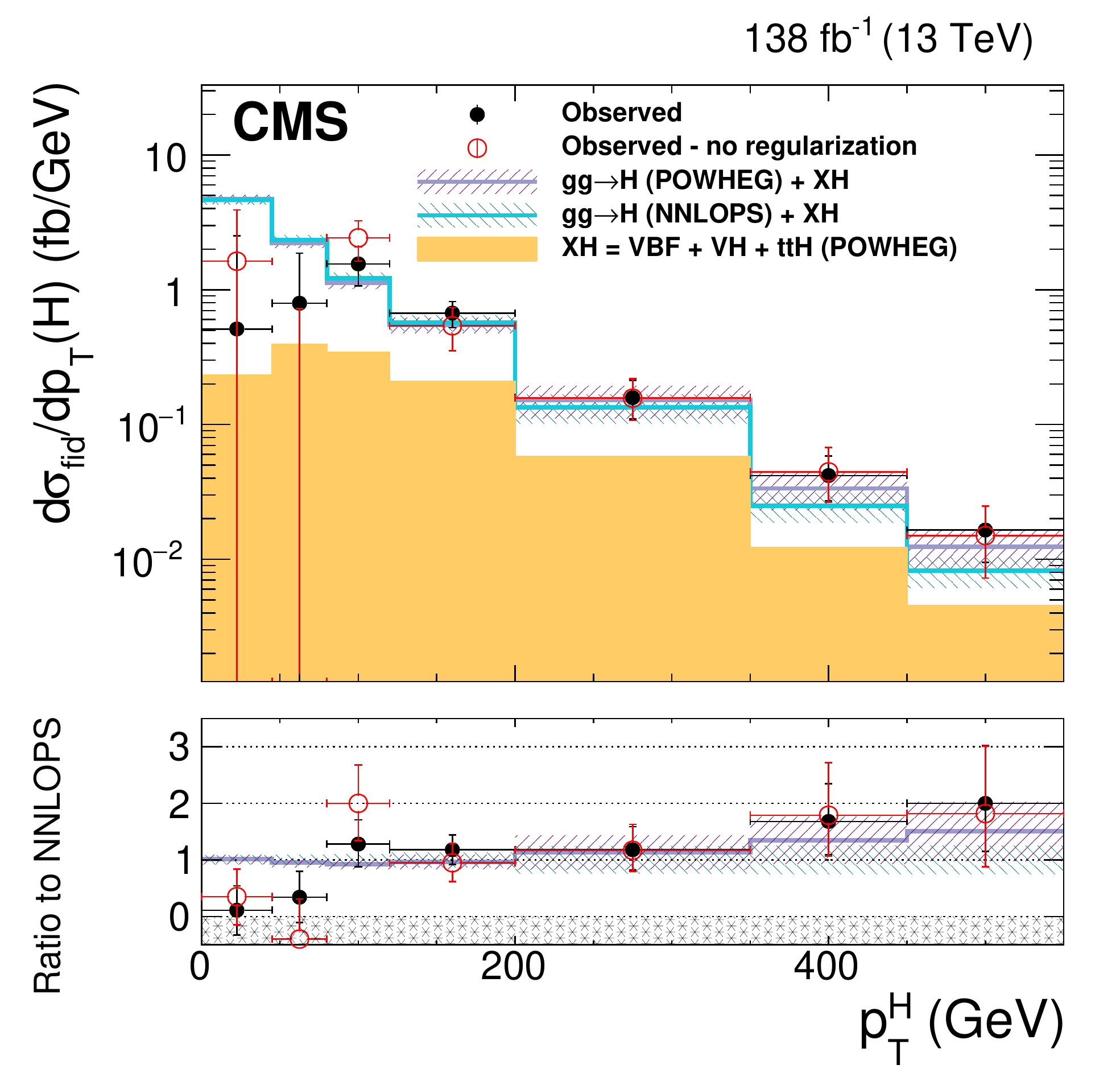}
  \includegraphics[width=0.4\textwidth]{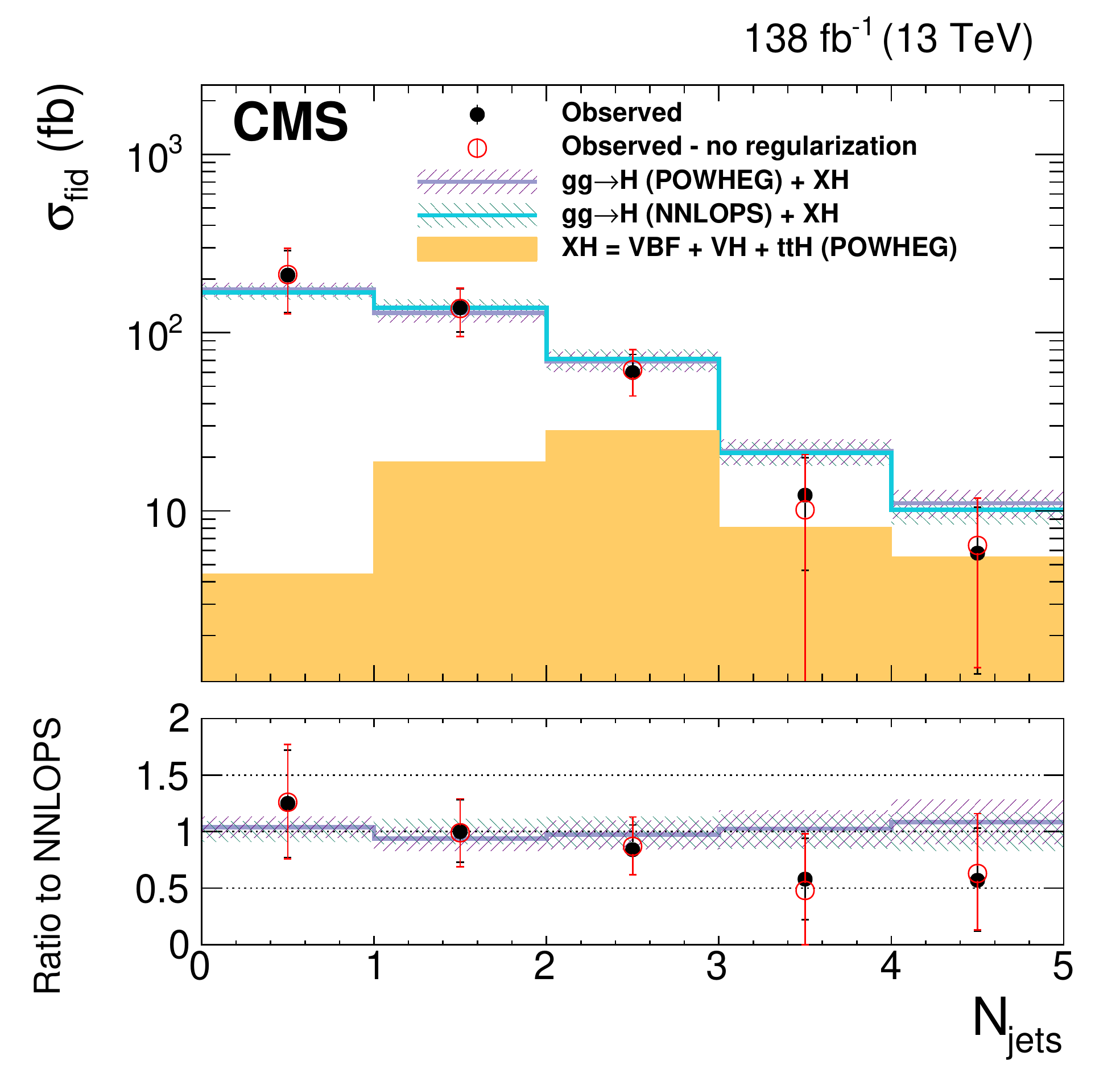}\\
  \includegraphics[width=0.4\textwidth]{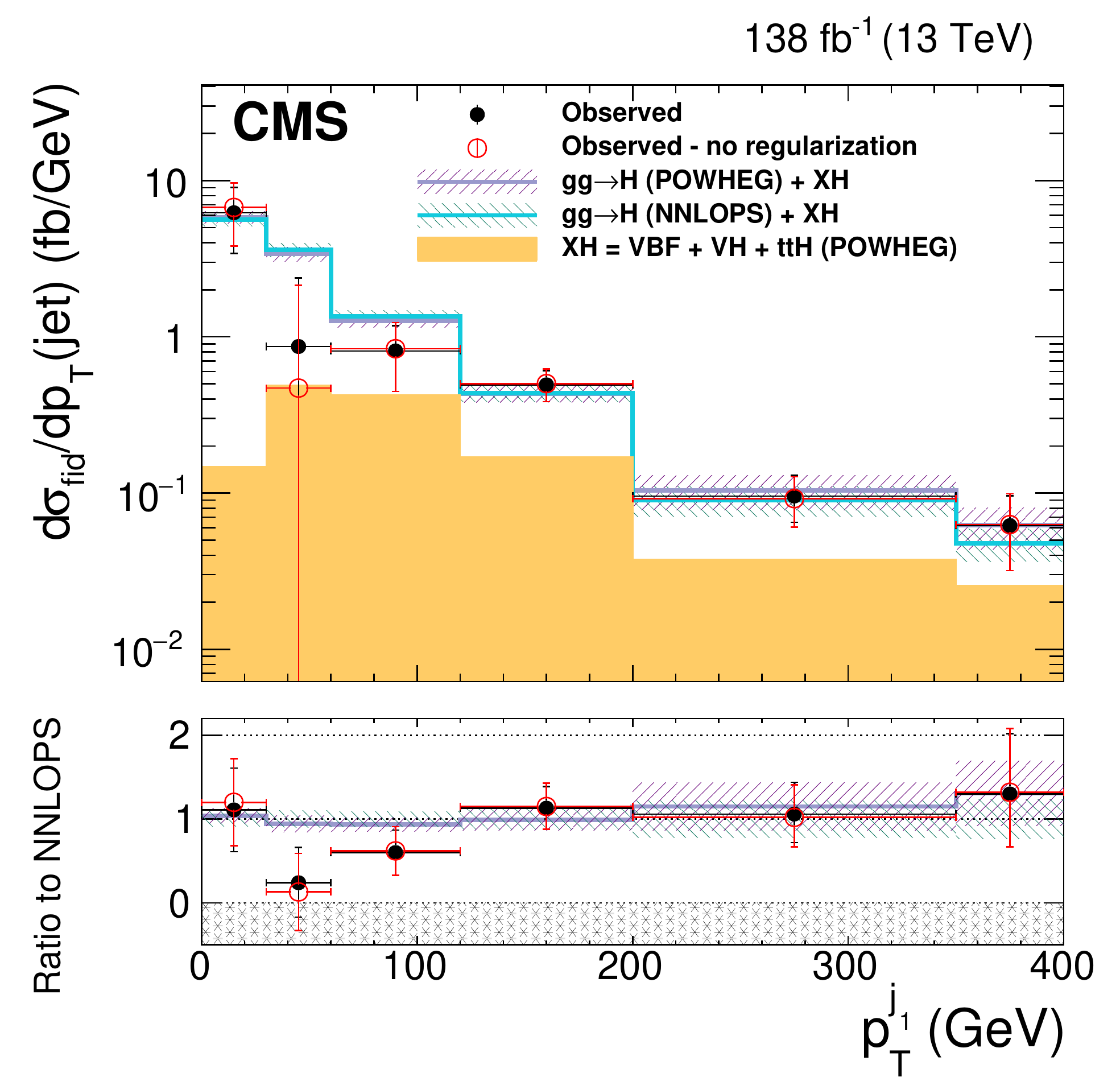}
  \caption{\label{fig:differential_hpt} Observed and expected differential fiducial cross section in bins of $\pth$ (\cmsLeft), $\njets$ (\cmsMiddle), and $\ljpt$ (\cmsRight). Both regularized (full markers) and unregularized (hollow markers) are shown. The most-left bin in the $\ljpt$ distribution includes all events without a jet with $\pt>30\GeV$. The uncertainty bands in the theoretical predictions include uncertainties from the following sources: PDF, renormalization and factorization scale, underlying event and parton showering, and branching fraction of the Higgs boson to \PGt leptons. The last bins include the overflow. }
\end{figure}

The predicted and measured differential fiducial cross sections are shown in Fig.~\ref{fig:differential_hpt} for the regularized fits. 
Tabulated results are available in the HepData database~\cite{hepdata} for the regularized and unregularized cases. 
The fit has a $p$-value 
with respect to the SM expectation from the $\textsc{nnlops}$ prediction of 17, 71, and 45\% for the measurements of $\pth$, $\njets$, and $\ljpt$, respectively. No significant deviation with respect to the SM predictions is observed, and the measurements are compatible with both the \POWHEG and $\textsc{nnlops}$ expectations. 
The low measured cross sections for $0<\pth<45\GeV$ and $45<\pth<80\GeV$ do not coincide with the much more precise 
measurements performed in this phase space in other final states~\cite{Sirunyan:2018sgc,ATLAS:2020wny}, and are attributed to statistical fluctuations.

The results are dominated by statistical and theoretical uncertainties.
After the maximum likelihood fit described later in this Letter, the uncertainty in the background with jets misidentified as $\tauh$ candidates is at the percent level in the 
phase space region with large background contributions, and up to 10--15\% at high $\pth$.
The impact on the template normalization from the uncertainties for embedded events without any reconstructed jet are 7 and 4\% in the case of no jets and one jet, respectively, and becomes negligible at high jet multiplicity. 
Acceptance uncertainties for the $\Pg\Pg\PH$ signal give the largest contribution to the overall impacts on the fit results from the theoretical part. The impacts on the fits from the uncertainties due to migration between different jet multiplicity bins are less than 8\% overall, 
while the combined effect of the other theoretical uncertainties is less than 3\%.

The measurement is precise with respect to the measurements in other 
final states for $120<\pth<600\GeV$, $\njets\geq2$, and $\ljpt>120\GeV$. More specifically, this measurement for $120<\pth<200\GeV$ 
is comparable in precision with the measurements by the CMS~\cite{CMS:2021ugl} and ATLAS~\cite{ATLAS:2020wny} Collaborations in the $\PH\to\PZ\PZ\to 4\ell$ decay channel with 137--139\fbinv,  
and 50\% more sensitive than the CMS measurement in the $\PH\to\PW\PW$ channel with 137\fbinv~\cite{Sirunyan:2020tzo} and 
the combination performed by the CMS Collaboration with 36\fbinv in the $\PQb\PQb$, $\PGg\PGg$, and $\PZ\PZ$ decay channels~\cite{Sirunyan:2018sgc}. 
For $200<\pth<600\GeV$, the current measurement 
has a significantly higher precision and granularity 
than the measurements in Refs.~\cite{Khachatryan:2016vnn,Aaboud:2018ezd,Sirunyan:2018sgc,Sirunyan:2018kta,Sirunyan:2020tzo,ATLAS:2020wny,CMS:2021ugl}. 

The inclusive fiducial cross section
is measured from the distributions used in the differential measurements of $\njets$, by reformulating the 
parameters of interest such that one modifies the total inclusive fiducial cross section. 
Its measured value is $426\pm 102\unit{fb}$, compatible with the SM expectation of $408\pm 27\unit{fb}$.

In summary, measurements of the differential fiducial cross sections of the Higgs boson have been 
performed for the first time at the LHC in the decay channel of two \PGt leptons. 
The differential cross sections as functions of the Higgs boson transverse momentum, 
the jet multiplicity, and transverse momentum of the leading jet, 
are in agreement with the expectations of the standard model, with a competitive precision with respect to 
measurements in other final states in the phase spaces with a large jet multiplicity, or with a Higgs boson transverse 
momentum above 120\GeV. 
In addition, the fiducial inclusive
cross section has been measured to be $426\pm 102\unit{fb}$, in agreement with the standard-model expectation of $408\pm 27\unit{fb}$.

\begin{acknowledgments}
  We congratulate our colleagues in the CERN accelerator departments for the excellent performance of the LHC and thank the technical and administrative staffs at CERN and at other CMS institutes for their contributions to the success of the CMS effort. In addition, we gratefully acknowledge the computing centers and personnel of the Worldwide LHC Computing Grid and other centers for delivering so effectively the computing infrastructure essential to our analyses. Finally, we acknowledge the enduring support for the construction and operation of the LHC, the CMS detector, and the supporting computing infrastructure provided by the following funding agencies: BMBWF and FWF (Austria); FNRS and FWO (Belgium); CNPq, CAPES, FAPERJ, FAPERGS, and FAPESP (Brazil); MES (Bulgaria); CERN; CAS, MoST, and NSFC (China); MINCIENCIAS (Colombia); MSES and CSF (Croatia); RIF (Cyprus); SENESCYT (Ecuador); MoER, ERC PUT and ERDF (Estonia); Academy of Finland, MEC, and HIP (Finland); CEA and CNRS/IN2P3 (France); BMBF, DFG, and HGF (Germany); GSRT (Greece); NKFIA (Hungary); DAE and DST (India); IPM (Iran); SFI (Ireland); INFN (Italy); MSIP and NRF (Republic of Korea); MES (Latvia); LAS (Lithuania); MOE and UM (Malaysia); BUAP, CINVESTAV, CONACYT, LNS, SEP, and UASLP-FAI (Mexico); MOS (Montenegro); MBIE (New Zealand); PAEC (Pakistan); MSHE and NSC (Poland); FCT (Portugal); JINR (Dubna); MON, RosAtom, RAS, RFBR, and NRC KI (Russia); MESTD (Serbia); SEIDI, CPAN, PCTI, and FEDER (Spain); MOSTR (Sri Lanka); Swiss Funding Agencies (Switzerland); MST (Taipei); ThEPCenter, IPST, STAR, and NSTDA (Thailand); TUBITAK and TAEK (Turkey); NASU (Ukraine); STFC (United Kingdom); DOE and NSF (USA).
\end{acknowledgments}

\bibliography{auto_generated}

\cleardoublepage \appendix\section{The CMS Collaboration \label{app:collab}}\begin{sloppypar}\hyphenpenalty=5000\widowpenalty=500\clubpenalty=5000\vskip\cmsinstskip
\textbf{Yerevan Physics Institute, Yerevan, Armenia}\\*[0pt]
A.~Tumasyan
\vskip\cmsinstskip
\textbf{Institut f\"{u}r Hochenergiephysik, Wien, Austria}\\*[0pt]
W.~Adam, J.W.~Andrejkovic, T.~Bergauer, S.~Chatterjee, M.~Dragicevic, A.~Escalante~Del~Valle, R.~Fr\"{u}hwirth\cmsAuthorMark{1}, M.~Jeitler\cmsAuthorMark{1}, N.~Krammer, L.~Lechner, D.~Liko, I.~Mikulec, P.~Paulitsch, F.M.~Pitters, J.~Schieck\cmsAuthorMark{1}, R.~Sch\"{o}fbeck, D.~Schwarz, S.~Templ, W.~Waltenberger, C.-E.~Wulz\cmsAuthorMark{1}
% ...
\vskip\cmsinstskip
% ...
1:  Also at TU Wien, Wien, Austria\\
% ...
\end{sloppypar}
\end{document}